# 500 GHz plasmonic Mach-Zehnder modulator enabling sub-THz microwave photonics

Maurizio Burla,[1,*] Claudia Hoessbacher,[1] Wolfgang Heni,[1] Christian Haffner,[1] Yuriy Fedoryshyn,[1] Dominik Werner,[1] Tatsuhiko Watanabe,[1] Hermann Massler,[2] Delwin Elder,[3] Larry Dalton,[3] and Juerg Leuthold[1]

[1] *Institute of Electromagnetic Fields, ETH Zurich, Gloriastrasse 35, Zurich 8092, Switzerland*
[2] *Fraunhofer IAF, Tullastraße 72, 79108 Freiburg im Breisgau, Germany*
[3] *Department of Chemistry, University of Washington, Seattle, WA 98195-1700, United States*
\* *Corresponding author:* maurizio.burla@ief.ee.ethz.ch

**Abstract:** Broadband electro-optic intensity modulators are essential to convert electrical signals to the optical domain. The growing interest in THz wireless applications demands modulators with frequency responses to the sub-THz range, high power handling and very low nonlinear distortions, simultaneously. However, a modulator with all those characteristics has not been demonstrated to date. Here we experimentally demonstrate that plasmonic modulators do not trade off any performance parameter, featuring – at the same time – a short length of 10s of micrometers, record-high flat frequency response beyond 500 GHz, high power handling and high linearity, and we use them to create a sub-THz radio-over-fiber analog optical link. These devices have the potential to become a new tool in the general field of microwave photonics, making the sub-THz range accessible to e.g. 5G wireless communications, antenna remoting, IoT, sensing, and more.

## 1. Introduction

Microwave photonic applications in the THz range are recently attracting a growing interest, due to the need to find solutions for next-generation (5G) wireless communication systems capable of unprecedented data rates. Over the last few years, in fact, we have seen an explosive growth of wireless data traffic, driven by widespread adoption of high-bandwidth services on mobile devices[1]. To enable the expected transmission rates of tens or even hundreds of Gb/s, without resorting to very high order modulation formats, carrier frequencies in the unallocated regions of the electromagnetic spectrum above 300 GHz are required[1-5], Fig. 1(a). THz signals above 300 GHz can be generated and detected either using all-electronic devices or via photonic techniques. Since the backbone of the internet is composed by very high-capacity fiber optic cables, the photonic approach has the ultimate advantage to allow seamless integration with the existing fiber networks[4]. This, however, requires optical-to-THz and THz-to-optical converters with bandwidth well above 300 GHz, high power handling and high linearity. While uni-travelling carrier photodiodes (UTC-PD) are an established solution for optical-to-THz conversion[5], including integration with sub-THz waveguides[6], efficient THz-to-optical conversion is still a challenge for the radio-over-fiber community[7, 8], as it requires modulators with electro-optic bandwidth well above 300 GHz, high power handling and very high linearity[9], which have not been demonstrated to date in any single device. In addition, besides THz communications, such modulators would potentially enable many more applications to reach the THz frequency range, such as high-speed airborne mm-wave photonic relays[10], agile mm-wave communication using photonic frequency conversion[11], signal distribution for mm-wave satellite instrumentation[12], photonic signal processing of radio signals[13] or mm-wave radar[14, 15].

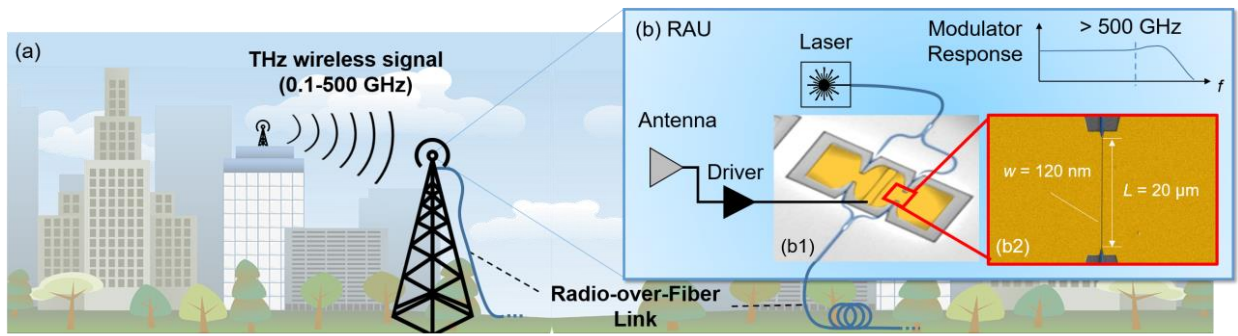

Fig. 1. (a) A THz wireless communication scenario. (b) At the remote antenna unit (RAU), THz wireless signals are received by an antenna and converted to the optical domain to be transported over a radio-over-fiber analog link. A modulator with sub-THz bandwidth, high linearity and high-power handling is needed to encode the THz signal onto an optical carrier with high fidelity. Inset: (b1) micrograph of a plasmonic Mach-Zehnder modulator with a plasmonic phase modulator in each arm; (b2) image of the 120 nm-wide, 20 µm-long phase modulator slot waveguide.

Current modulators optimized for analog applications are mostly based on lithium niobate[16], gallium arsenide[17], indium phosphide[18] and, more recently, silicon[19]. Those devices have shown good performance in terms of linearity and power handling; however, commercial solutions require large mm- or even cm-scale travelling wave structures for broadband performance and are limited to speeds below 110 GHz. Over the last four years, a novel class of modulators, known as plasmonic modulators, has been introduced and demonstrated ultra-compact footprints[20] (10s $\mu m^2$), ultra-low power consumption (2.8 fJ/bit at 100 GBd)[21], and flat frequency responses up to 170 GHz[22] and 325 GHz[23]. Interestingly, in most recent structures reasonably low loss were reported (2.5 dB in-chip losses for a ring modulator[20] and 8 dB for a Mach-Zehnder modulator[24]). Very recently, plasmonics has shown the ability to detect THz pulses using bow-tie shaped structures[25]. However, while all of these results are of highest interest, a modulator simultaneously displaying sub-THz frequency responses, high power handling and high linearity has not yet been shown.

Here we experimentally demonstrate plasmonic Mach-Zehnder modulators (MZM) with a flat frequency response up to 500 GHz. The modulators are below 25 micrometers in length and display high linearity, in line with those of high-performance commercial modulators optimized for analog applications. We use them to demonstrate an analog radio-over-fiber (RoF) link up to 325 GHz, with >100 GHz bandwidth, only limited by our electrical measurement equipment. This indicates that plasmonic modulators have a strong potential not only for digital communications, but they can also handle the stringent requirements needed for high-performance microwave photonics applications, communications, sensing, and more.

## 2. 500 GHz Mach-Zehnder modulator

Plasmonic modulators are based on the concept of surface plasmon polaritons (SPPs). SPPs are electromagnetic surface waves propagating at dielectric-metal interfaces[26-28]. For our modulators we resort to the plasmonic organic hybrid (POH) platform[29]. A plasmonic phase modulator is depicted Fig. 1(b2). It consists of two metallic electrodes forming a metal-insulator-metal (MIM) slot waveguide. The slot is filled with a nonlinear organic material[30, 31] whose refractive index changes linearly with the applied electric field, according to the Pockels effect[32]. A silicon strip waveguides (450 nm-wide, 220 nm-thick) feeds light to the plasmonic slot waveguide. A linear taper transforms the photonic mode in the silicon nanowire to a SPP that propagates along the slot, where its phase is modulated by the applied field, and is then coupled back to the photonic waveguide by an identical output taper. High modulation indices are obtained thanks to the strong electrical and optical field confinement in the slot, and an excellent overlap of the two fields. Thanks to the high modulation efficiency, the plasmonic slot can be kept short and, in turn, optical losses are reduced. A very important advantage of this kind of modulators for MWP applications is their ultra-broad electro-optical bandwidth, due to the quasi-instantaneous nonlinear effect and the extremely small *RC* time constant of the structure. In fact, the small slot height and length reduce its capacitance to the femtofarad range, while the resistance is low (sub-1 Ω, neglecting the contact pads) thanks to the high conductivity of the metal electrodes. This yields a theoretical cutoff frequency in the THz range, according to theoretical predictions [33]. This, in turn, also enables operation in the mm-wave and sub-THz range as desired for MWP applications.

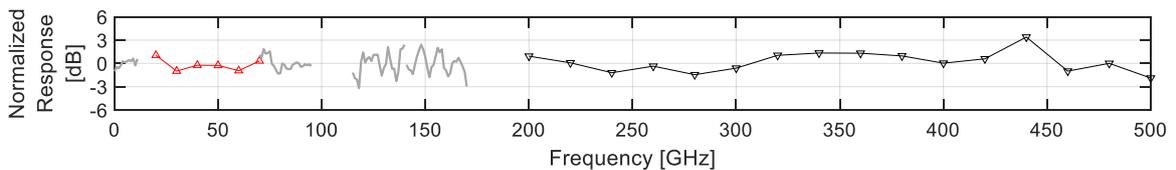

Fig. 2. Measured response of the POH-MZM output for input frequency from 75 MHz to 500 GHz. The figure displays the measured power of the optical modulation sidebands, normalized to the electrical input power to the POH-MZM. The device features a flat response up to 500 GHz.

The device employed in the test is a POH-MZM[34], Fig. 1(b1), composed by a silicon strip-waveguide interferometer with one 20 µm–long POH phase modulator in each arm, Fig. 1(b2). In this implementation, each phase modulator is realized with a ~120 nm-wide slot waveguide filled with the organic electro-optic material composite HD-BB-OH/YLD124[31]. This type of composites have shown electro-optic coefficients in POH devices reaching 325 pm/V when operating close to the resonance wavelength[35]. The phase modulators are driven in push-pull using a ground-signal-ground (GSG) electrical probe configuration. An extinction ratio of approximately 25 dB has achieved, indicating an accurate amplitude balance in the two arms of the Mach-Zehnder interferometer. Cut-back measurements show a loss for the un-optimized optical grating couplers of approximately 5 dB/coupler. Each photonic-plasmonic converter[33] display a loss of approximately 1.1 dB. This allows to estimate a loss in the plasmonic slot in the order of 0.5 dB/µm.

In the next step, we describe the experiments that have been performed to assess the frequency response of the POH-MZM to electrical signals from 75 MHz up to 500 GHz onto an optical continuous-wave (CW) carrier. In the experiments, a CW laser at 1547.5 nm with 0 dBm optical power was fed into the POH-MZM. The frequency response to the electrical signals were then determined in two different setups that covered the frequency ranges from 15 GHz – 70 GHz and 200 GHz – 500 GHz. Subsequently, the optical spectrum of the intensity-modulated carrier at the modulator output was measured using an optical spectrum analyzer (OSA). The amplitude ratio between the fundamental and first side lobe in the spectrum was then used to determine the relative response of the electrical signal onto the optical signal[36]. The frequency response was then normalized with respect to the power of the electrical signal driving the modulator.

The results of the frequency response measurements are depicted in Fig. 2. The figure shows a flat frequency response from 75 MHz to 500 GHz, compiled from the two different setups as above. The response is flat with a ripple always below ±3 dB (the origin of the ripple is discussed in the *Supporting Information*). As a consequence, we are not yet able to measure a bandwidth limitation since the response never crosses the -3 dB line compared to the low-frequency response. Additionally, we include (grey solid lines) the measurements from 75 MHz to 15 GHz, from 70 GHz to 95 GHz and from 115 GHz to 170 GHz, as reported by Hoessbacher *et al.*[22], who obtained them with a similar POH-MZM device. Both measurements have been normalized to their low frequency value to show that this kind of modulators feature a flat frequency response over the complete measurable spectral region, and show no signs of bandwidth limitation. Frequency ranges from 95 GHz to 115 GHz and from 170 GHz to 200 GHz cannot be covered in our laboratory.

Below 170 GHz, the test signal was generated by electrical means, using a signal generator and appropriate frequency multipliers. In the frequency range between 200 and 500 GHz, the test signal was generated by optical means via optical heterodyning, using two tunable laser sources and a UTC-PD to generate a THz signal at the frequency difference between the two lasers. The UTC-PD displays a 3 dB bandwidth in the 270 GHz – 370 GHz range; however, it is still capable of generating signals down to 200 GHz and up to about 500 GHz. Full details on the experimental setups and the procedure used to measure the modulator response are reported in the *Supporting Information*.

## 3. Linearity and Power Handling

In analog photonic applications, linearity is a crucial parameter as it directly affects the system spurious-free dynamic range (SFDR)[9, 37]. The SFDR is a figure of merit that indicates the realistic power range over which a given network is capable of operating in a linear regime without being limited by noise[9]. For an intensity modulator, linear operation indicates that the modulating signal is transferred with high fidelity to the intensity of the light, i.e. without introducing any spurious frequency components. Deviations from the linear regime, e.g. generation of spurious tones, are known as nonlinear distortions. Those are important since, once generated, they normally can no longer be removed. In the following we report the experimental characterization of the nonlinear distortions introduced by a POH-MZM. The most common way to probe these nonlinear effects is the use of the method known as two-tone test[37]. This method allows us to characterize the second- and third-order intermodulation distortions (IMD2 and IMD3), which limit the SFDR in sub-octave and multi-octave conditions, respectively[37]. Importantly, however, the direct measurement of the SFDR value does not allow to immediately draw conclusions on the linearity of the modulator alone, because it includes the effects of noise and loss terms in the link. Therefore, in order to analyse nonlinear distortions originating from the modulator, we measure the IMD2 and IMD3 terms, as suggested in literature[37]. Through this test, we could experimentally compare the linearity of the POH-MZM with the one of a commercial GaAs modulator optimized for high-performance analog applications. The results in Fig. 3 show that the third-order intermodulation distortions from the POH-MZM are as low as those obtained from the commercial modulator.

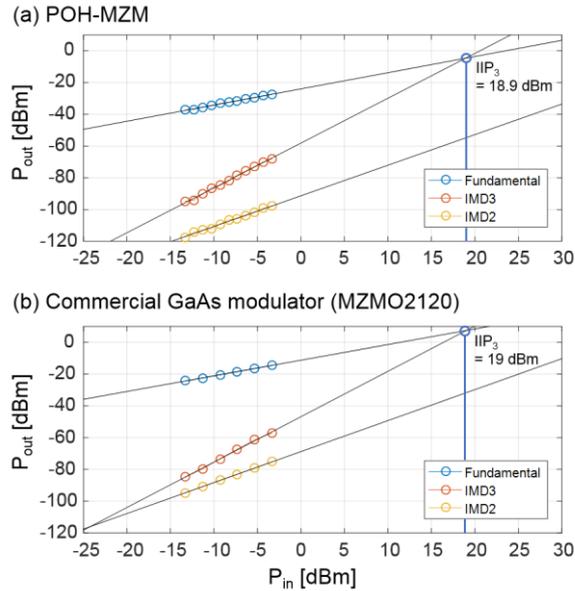

Fig. 3. Second-order (IMD2) and third-order (IMD3) intermodulation distortions for the POH-MZM (a) and a commercial GaAs modulator (b). The third-order input intercept point for the POH-MZM is at 18.9 dBm, very close to the 19.0 dBm obtained with a commercial high-performance GaAs modulator of similar half-wave voltage.

Experimental two-tone tests are performed on POH-MZM device with phase modulators having a slot length of 25 µm and a width of approximately 65 nm. Tests were performed at a frequency of 21 GHz ± 1 kHz (only limited by the available microwave components and test equipment), when the modulator is operated in its quadrature point, as discussed above, using a high power handling photodiode (PD). The power at the PD input was kept at least 5 dB below its maximum to ensure that distortions originating from the PD are negligible. The power of the input tones is swept between -13.3 dBm and -3.3 dBm (corresponding to driving voltages ranging from 274 mV peak-to-peak to 865 mV peak-to-peak on this load). The corresponding powers of the fundamental ($f_1$, $f_2$), IMD2 ($f_1+f_2$) and IMD3 ($2f_1-f_2$ and $2f_2-f_1$) tones are measured with an electrical spectrum analyzer (ESA) and reported in Fig. 3. Details on the experimental setup are reported in the *Supporting Information*.

The intercept point of the extrapolation of the fundamental and the IMD3 curve is known as third-order intercept point (IP3). The input power corresponding to this value (3rd order input intercept point, IIP3) amounts to 18.9 dBm. For comparison, we have measured a commercial GaAs modulator ($V_\pi$ = 3.0 V at 20 Gb/s PRBS) using the same setup and the same average optical power impinging on the PD. The measured IIP3 in quadrature was 19.0 dBm. This result demonstrates that the linearity distortions originating from third-order intermodulations in POH-MZM are at least comparable to those obtained using commercial high-integrity modulators optimized for both telecom and radio-over-fiber applications.

Using current un-optimized grating couplers (~ 5 dB loss each) leads to a link gain in the order of -24 dB. This can be increased by reducing the optical propagation losses and the fiber-chip coupling losses, which could be improved e.g. using more efficient grating couplers or edge coupling. We expect that an RF gain improvement of about 10 dB or more can be achieved by reducing coupling loss to state-of-the-art values of approximately 2.5 dB per grating.

Using the same setup, high RF power tests have also been performed. We employed two high power RF amplifiers, to feed two signals simultaneously at 1 GHz. The modulator showed operation without damage up to 21.4 dBm (~138 mW, approx. 14.9 V peak-to-peak) of electrical input power for each tone - only limited by the power of the available RF amplifiers. We believe that the favorable higher power handling without any damage is due to the fact that - since the modulators are very short - we operate them by capacitive loading, where only little energy is dissipated in the capacitor itself. Also, an efficient heat dissipation is provided by the high thermal conductivity of the gold electrodes constituting the modulator slot. Further tests have also shown that the modulator can handle high optical input power. Non-destructive tests showed normal operation for on-chip optical input power up to at least 16 dBm. Taking into account the grating coupler losses provided above, the optical power impinging on the modulator input amounts to approximately 11 dBm.

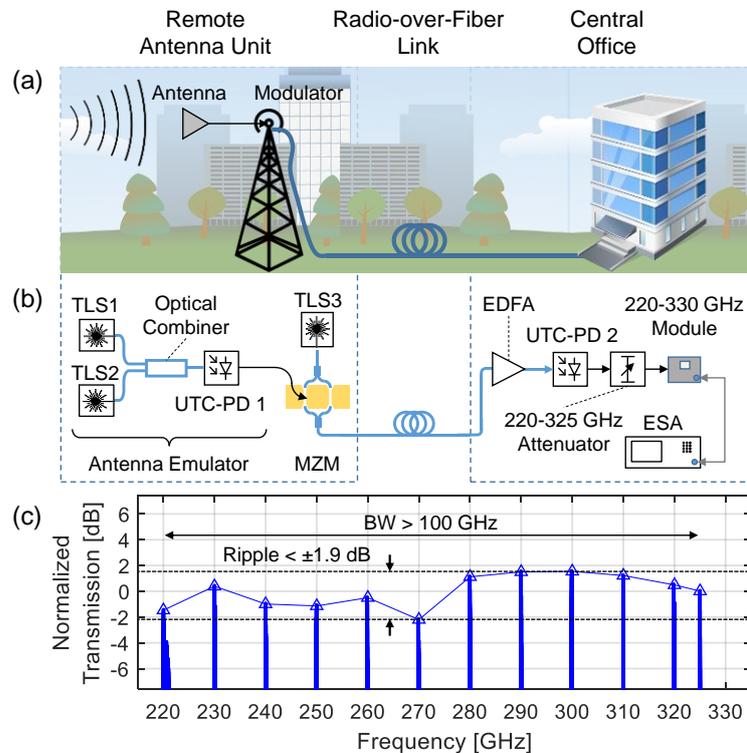

Fig. 4. (a) RoF link scenario connecting a remote antenna to a central office. (b) Experimental setup. (c) Normalized transmission of the analog radio-over-fiber link. The link operates between 220 and 325 GHz, only limited by the available electrical test equipment.

## 4. 325 GHz Analog Radio-over-Fiber Link

To further prove the capability of the modulator in Fig. 1, we use it to implement an analog RoF link, as depicted in Fig. 4(a), demonstrating direct THz-to-optical-to-THz conversion over >100 GHz bandwidth (220-325 GHz range), only limited by our electrical test equipment. Fig. 4(b) shows the setup and Fig. 4(c) displays the broadband RoF link transmission response, normalized to the UTC-PD responsivity. As done for the modulator response test, sub-THz waves are generated using UTC-PD 1, emulating an antenna source, and are used to directly modulate an optical carrier using a POH-MZM, Fig. 4(b). The modulated signal propagates through a radio-over-fiber link. The mm-waves are then detected with another UTC-PD and analysed using an ESA with a mm-wave extension module. Further details on the setup and the calibration procedure are reported in the *Supporting Information*. This experiment demonstrates the modulator capability to perform direct THz-to-optical conversion and confirms its flat response. Residual ±1.9 dB variations across the band are attributed to ripples in the attenuator loss, which could not be tested in our laboratory.

## 5. Conclusions

We have experimentally demonstrated Mach-Zehnder modulators simultaneously meeting all requirements for analog applications, i.e. high linearity, high power handling and speeds reaching 500 GHz. To our knowledge, we reported the fastest Mach-Zehnder modulator to date. We also demonstrate a radio-over-fiber link up to 325 GHz (>100 GHz bandwidth). These results suggest that plasmonics has the potential for becoming a new tool to the field of microwave photonics, enabling applications to reach the sub-THz range, while preserving the high-performance required in large-scale analog applications such as 5G wireless, antenna remoting, IoT, sensing, and more.

## 6. Funding

SNSF Ambizione Grant (173996); ERC PLASILOR project (670478); H2020 PLASMOfab project (688166); NSF (DMR-1303080); AFOSR (FA9550-15-1-0319).

## 7. Acknowledgment

The authors wish to thank prof. C. Bolognesi, H. Benedikter and M. Leich for support with the experiments, and NTT for providing the UTC-PD. M. Burla acknowledges Dr. D. Marpaung for the fruitful discussions.

# 500 GHz plasmonic Mach-Zehnder modulator enabling sub-THz microwave photonics: Supporting Information

Maurizio Burla,[1,*] Claudia Hoessbacher,[1] Wolfgang Heni,[1,] Christian Haffner,[1] Yuriy Fedoryshyn,[1] Dominik Werner,[1] Tatsuhiko Watanabe,[1] Hermann Massler,[2] Delwin Elder,[3] Larry Dalton,[3] and Juerg Leuthold[1]

[1] Institute of Electromagnetic Fields, ETH Zurich, Gloriastrasse 35, Zurich 8092, Switzerland
[2] Fraunhofer IAF, Tullastraße 72, 79108 Freiburg im Breisgau, Germany
[3] Department of Chemistry, University of Washington, Seattle, WA 98195-1700, United States
* Corresponding author: maurizio.burla@ief.ee.ethz.ch

This document provides supporting information to "500 GHz plasmonic Mach-Zehnder modulator enabling sub-THz microwave photonics". We describe details on the experimental setups employed to measure the frequency response of the modulator, over two different frequency ranges (from 20 to 70 GHz and from 200 to 500 GHz), and the procedure used to extract the modulator response. We describe the setup and procedure employed to measure the third-order intermodulation distortions via two-tone tests. Finally, we report details on the radio-over-fiber link experiment in the 220-325 GHz window.

## 1. Characterization of the modulator response: experimental setups

In this section we describe the experimental setups employed for the measurement of the modulator response over the frequency ranges 20-70 GHz and 200-500 GHz. Two different setups were used (Fig. S1).

**Frequency range 20-70 GHz, Fig. S1(a).** Below 70 GHz, an RF signal generator (Synthesizer, Agilent E8257D) was used to directly generate the RF signal, which was fed to the modulator electrodes using a coaxial cable and a 67 GHz microwave probe. A vector network analyzer was used to characterize the loss of the RF cable versus frequency.

Setup for 20 GHz - 70 GHz

Setup for 200 GHz - 500 GHz

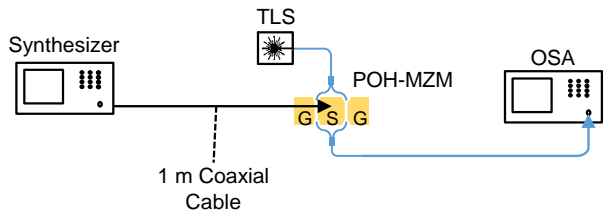
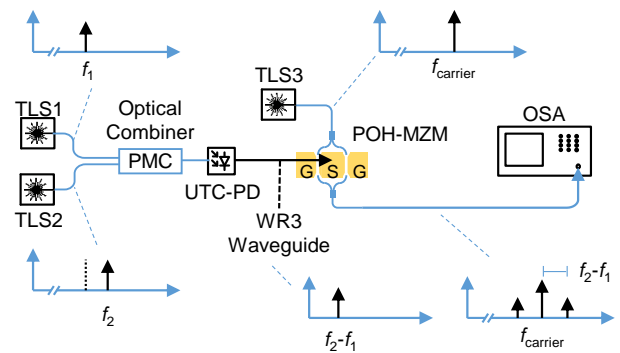

Fig. S1. Setup employed for the modulator bandwidth tests, over the frequency ranges: (a) 20-70 GHz; (b) 200-500 GHz. TLS: tunable laser source; POH-MZM: plasmonic-organic-hybrid Mach-Zehnder modulator; PMC: polarization-maintaining coupler; UTC-PD: uni-travelling-carrier photodetector ; OSA: optical spectrum analyzer.

**Frequency range 200-500 GHz, Fig. S1(b).** The experimental setup employed for the high-speed electrical tests in the 200-500 GHz range is displayed in Fig. S1(b). In this case, the sub-THz signals were generated optically. Two tunable laser sources (TLS1 and TLS2) generate optical carriers at $f_1$ = 191.5 THz and $f_2$ = 191.8 THz, respectively. The signals are combined using a polarization-maintaining coupler (PMC) and are fed to a uni-travelling carrier photodiode (UTC-PD) by means of a polarization maintaining fiber (PMF), where they induce the generation of a sub-THz signal via optical heterodyning. The generated THz wave has a frequency equal to the frequency difference of the two optical carriers such as $f_2$-$f_1$ = 300 GHz for the example above. The sub-THz signal is then coupled to a WR3 waveguide and fed to the modulator using a RF wafer probe with WR3 waveguide input and ground-signal-ground (GSG) configuration.

Setups suitable for testing the response in the 75 MHz to 15 GHz, 70 GHz to 110 GHz and 115 GHz to 170 GHz ranges, described by Hoessbacher *et al.* [1], were not available at the time of the experiment.

**Frequency tuning.** In the range 20-70 GHz, the signal frequency is changed by directly changing the frequency of the synthesizer in Fig. S1(a). In the range 200-500 GHz, the signal frequency can be varied optically, by keeping the first laser, TLS1 in Fig. S1(b), at 191.5 THz, while the second laser, TLS2 in Fig. S1(b), is swept between 191.7 THz and 192.0 THz in steps of 20 GHz. As a consequence, the frequency of the sub-THz signal generated by the UTC-PD ranges from 200 GHz to 500 GHz.

It is important to note that the UTC-PD in Fig. S1(b) displays a 3 dB bandwidth in the 270 GHz – 370 GHz range; however, it is capable of generating signals down to 200 GHz and up to about 500 GHz. During the sweep, the optical power of the two lasers is kept constant at 16 dBm and the UTC-PD photocurrent is approximately 6.5 mA.

## 2. Extraction of the modulator frequency response

Fig. S2(a) shows the normalized modulator frequency response. Figs. S2(b-c) illustrate how the frequency response in Fig. S2(a) has been obtained. In detail: Fig. S2(b) displays the optical spectra measured at the modulator output for different frequencies and powers of the input electrical signal; Fig. S2(c) shows the measured optical sideband power (blue circles) extracted from Fig. S2(b) compared to the actual electrical power driving the modulator (red triangles), at different modulating frequencies. The ratio of these two quantities (black inverted triangles) gives the normalized modulator response in Fig. S2(a).

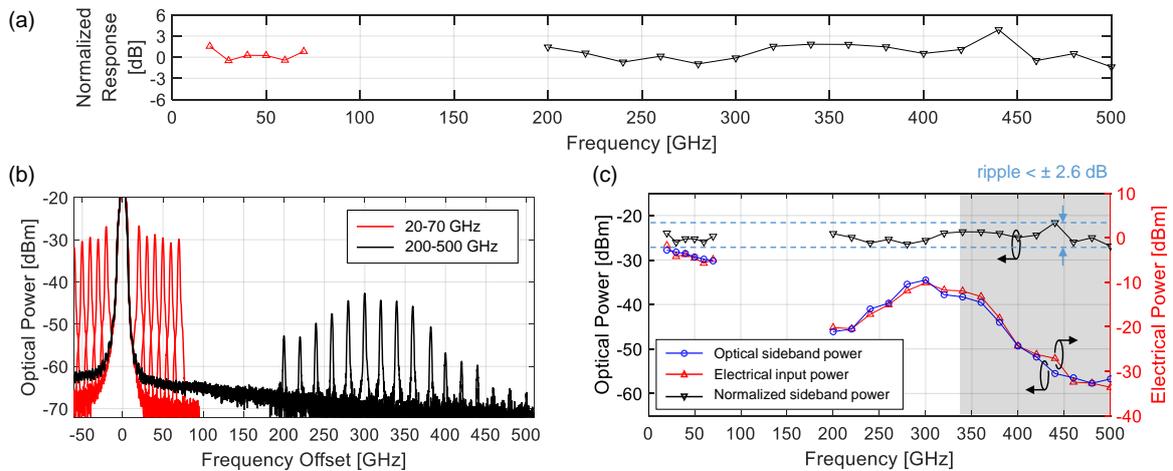

Fig. S2. (a) Measured optical spectrum at the POH-MZM output for input frequency in the ranges 20-70 GHz (red triangles) and 200-500 GHz (black inverted triangles). (b) Optical spectra measured using the optical spectrum analyzer (OSA). (c) Power of the optical sidebands (blue circles) compared to the electrical power driving the modulator (red triangles). The optical sideband power follows very closely the power level of the electrical signal feeding the modulator, indicating a flat modulator frequency response (black inverted triangles), with a ripple always below ±2.6 dB. The gray area indicates the frequency region above 347.143 GHz, where the WR3 waveguide is no longer single mode.

Note that the electrical power reaching the modulator varies largely over frequency, due to both the effect of the source and the effect of the RF probes and cables or waveguides. All these effects were experimentally characterized, as discussed below, and taken into account when normalizing the modulator response.

**Variation of the optical sideband power.** From Fig. S2(b) it can be noted that the measured sideband amplitude varies largely when sweeping the frequency. This is due to the fact that the UTC-PD, for constant optical input power, has very different responsivity at different frequencies. In addition, the WR3 waveguide probe also displays a rapid transmission roll-off above 360 GHz. The probe insertion loss has been experimentally characterized between 200 and 500 GHz with a dedicated vector network analyzer test set, using WR2.2 extension heads and WR2.2 to WR3 waveguide tapers.

**Comparison of optical sideband power and electrical power driving the modulator.** Fig. S2(c) shows the average power of the electrical signal that drives the POH-MZM contacts in comparison to the optical sideband power extracted from Fig. S2(b). By overlapping the two lines (average electrical power and measured optical sideband power), while keeping the same vertical scale (10 dB/division) shows that the optical sideband power follows very closely the power level of the electrical signal feeding the modulator, indicating a flat frequency response for the modulator over the whole frequency range. This can be better seen by accurately normalizing the measured optical sideband power to the UTC-PD output power profile and the probe transmission, as shown by the black line on top of Fig. S2(c).

**Discussion of the ripples in the normalized frequency response.** The normalized sideband power shows a ripple always below ±2.6 dB in the complete 200 GHz to 550 GHz range, and below ±1.4 dB in the 200-370 GHz span. In the highest frequency range, the ripple increases, while still keeping below ±2.6 dB. This can be explained considering that the WR3 waveguide (inner dimensions of 0.8636 mm × 0.4318 mm) is no longer single-mode above 347.143 GHz. As a consequence, above that frequency the power generated at the UTC-PD starts to be coupled to higher order modes, therefore making the power measurement more sensitive to experimental variations. The frequency range over which the waveguide is multimode is indicated by the grayed area in Fig. S2(c). Using UTC-PDs coupled to a waveguide probe with a higher cutoff frequency (e.g. WR2 or WR1) would allow not losing power to higher order modes. To our knowledge, these are not commercially available to date.

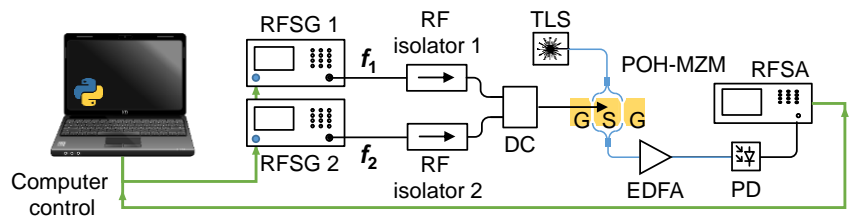

Fig. S3. Computer controlled experimental setup for nonlinear distortions tests. RFSG: RF signal generator; DC: directional coupler; TLS: tunable laser source; POH-MZM: plasmonic-organic-hybrid Mach-Zehnder modulator; EDFA: erbium-doped fiber amplifier; RFSA: RF spectrum analyzer (images from https://pixabay.com/).

## 3. Third-order intermodulation distortions: test setup

Fig. S3 shows the setup used for the nonlinear distortion tests [2, 3]. Two tunable RF signal generators (RFSG1 and RFSG2) produce two tones ($f_1$ and $f_2$) at 21 GHz ± 1 kHz, respectively. Each RFSG output is connected to an isolator (Ditom, 20-40 GHz). Signals at the isolator outputs are combined using a broadband 3 dB directional coupler (DC, Clear Microwave, 7-40 GHz). The signal at the coupler output drives the POH-MZM using a GSG probe. The optical carrier (+10 dBm) is provided by a tunable laser source (TLS). The light is carried on- and off-chip using silicon grating couplers [4], exactly like the frequency response test setup in Fig. S1. The optical output of the MZM is amplified using an erbium-doped fiber amplifier (EDFA) to an average optical power of 15 dBm and detected with a high-linearity, high power handling photodiode (PD, Optilab PD-20-HP-M, 20 dBm max. average input power). The power input to the PD was kept at least 5 dB below its maximum to ensure that distortions originating from the PD are negligible. The RF spectrum at the PD output is measured using an RF spectrum analyzer (RFSA). The POH-MZM is operated in quadrature [2] with zero bias by adjusting the wavelength of the TLS. Note that the test frequencies are only limited by the bandwidth of the available high-power photodetector (18 GHz at 3 dB).

The RF input power from the generators is swept from 0 dBm to 10 dBm. The attenuation of the RF paths between each RFSG and the POH-MZM input (including the effect of connectors, cables, solators, and combiners) amounts to 13.3 dB at 21 GHz. The corresponding powers of the fundamental ($f_1$, $f_2$), IMD2 ($f_1+f_2$) and IMD3 ($2f_1-f_2$ and $2f_2-f_1$) tones are measured with the RFSA and reported in Fig. 3 of the main paper.

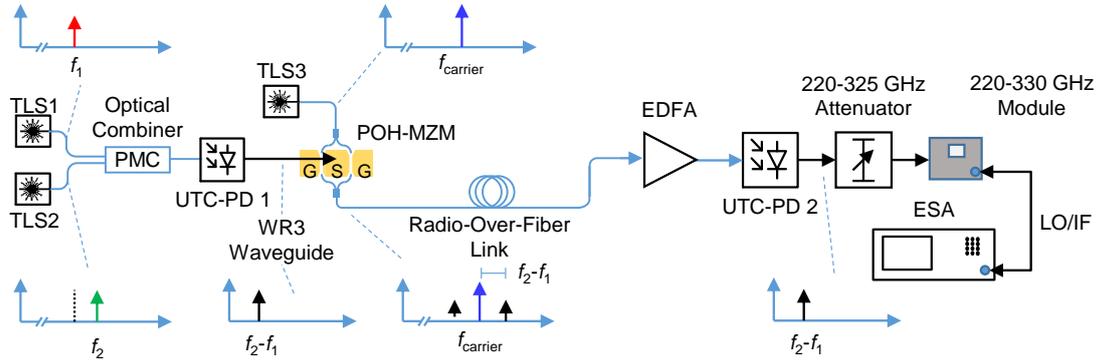

Fig. S4. Setup of the Radio-over-Fiber experiment at 220-325 GHz. The POH-MZM modulates an optical carrier $f_{carrier}$ with a mm-wave signal in the 220-325 GHz frequency range generated by UTC-PD 1. The modulated carrier is transported on a single-mode fiber (SMF) radio-over-fiber link, is amplified by an erbium-doped fiber amplifier (EDFA) and is received by UTC-PD 2. The detected mm-wave is analyzed with an electrical spectrum analyzer (ESA) using a mm-wave extension module.

### 4. Radio-over-fiber link experiment setup

We used the POH-MZM to implement a radio-over-fiber (RoF) link over the frequency range 220-325 GHz. Fig. S4 shows the experimental setup.

The mm-wave signal generation is performed via optical heterodyning, as in the frequency response test setup in Fig. S1(b). Two tuneable laser sources (TLS1 and TLS2), with an optical frequency difference varying between 220 GHz and 325 GHz, are combined with a polarization maintaining combiner and fed to a UTC-PD (UTC-PD 1), where they generate, by optical heterodyning, the THz wave to be transmitted on the RoF link. An optical carrier (generated by TLS3) enters the POH-MZM where it is intensity-modulated by the mm-waves generated by UTC-PD 1. At this point (differently from the frequency response test) the signal is propagated on a single-mode fiber (SMF) link and, after amplification by an erbium-doped fiber amplifier (EDFA), is received by another UTC-PD (UTC-PD 2). There, the signal is converted back from the optical to the mm-wave domain and is analysed by an electrical spectrum analyser (ESA, Agilent PXA N9030A), equipped with a 220-330 GHz signal analyser extension module (Keysight SAX247). A variable waveguide attenuator is added after UTC-PD 2 to protect the extension module. This attenuator element limits the analysis bandwidth to 220-325 GHz.

### 5. Calculation of the radio-over-fiber link response

The measured response of the RoF link is reported in the main paper, Fig. 4(c), after removing the effect of the test equipment used for the mm-wave spectrum analysis, and normalizing to the response of the receiver (UTC-PD 2). Fig. S5 shows how the curve in Fig. 4(c) is obtained.

The SAX247 220-330 GHz signal analyser extension module adds frequency-dependent loss terms that have to be removed from the raw measurement. Fig. S5(a) displays the specific loss contribution terms: the loss due to the PXA unit response when used with this extension module (red dashed line), and the mixer intrinsic loss (blue solid line) of the SAX247 extension module. Fig. S5(b) displays the RoF link response (where the previous effects have been removed) in comparison to the UTC-PD response (from the datasheet). For ease of comparison, both curves have been normalized to their average value, while keeping the same vertical scale. This shows that the RoF link has a transmission (inverted blue triangles) that grows by approximately 10 dB over the tested frequency range, up to approx.. 300-310 GHz, following the responsivity of the UTC-PD (red circles).

After removing the effect of the UTC-PD from the RoF response in Fig. S5(b), we obtain the line in Fig. 5(c) - or, equivalently, Fig. 4(c) in the main paper - which is relatively flat over frequency. This confirms the flatness of the POH-MZM response, as a further proof of its broadband operation. A residual variation of ±1.9 dB is visible across the band; this can be attributed to ripples in the attenuator loss response, which could not be measured with the network analyzers available in our laboratory.

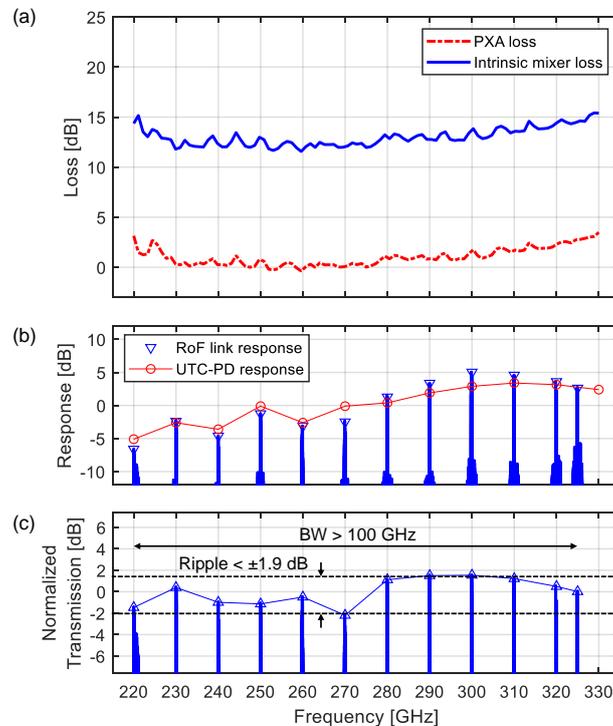

Fig. S5. Calculation of the normalized transmission of the RoF link in Fig. S4. (a) Loss contributions due to the measurement equipment (red dashed line: PXA unit loss; blue solid line: intrinsic mixer conversion loss in the SAX 247 unit). (b) Comparison of the measured spectrum of the RoF link with the UTC-PD responsivity, where the SAX 247 losses are removed. For ease of comparison, both curves are normalized to their mean value while keeping the same vertical scale. The comparison shows that the RoF response increases in frequency, following the responsivity of the UTC-PD, which is maximum around 310 GHz. (c) The difference between the two curves in Fig. S5(b) yields the normalized response of the RoF link, also reported in the main paper, Fig. 4(c). Residual deviations are attributed to the variations in band of the mm-wave attenuator loss, which could not be measured with our instrumentation.